\documentclass{article}
\usepackage{spconf,amsmath,graphicx}
\usepackage{amsfonts}
\usepackage{siunitx}
\usepackage{booktabs}
\usepackage{pgfplots,tikz}
\usetikzlibrary{positioning,fit,calc,shapes,matrix}
\usepackage{multirow}
\usepackage{cite}
\usepackage{url}
\usepackage{xcolor}
\usepackage{pifont}
\usepackage{etoolbox}

\def\L{{\cal L}}

\def\Ltriplet{{\L^{(\text{TRIPLET})}}}
\def\Lresmask{{\L^{(\text{res-mask})}}}
\def\Lmse{{\L^{(\text{MMSE})}}}

\def\vect#1{\mathbf{#1}}
\def\mat#1{\mathbf{#1}}

\title{Tackling real noisy reverberant meetings \\ with all-neural source separation, counting, and diarization system}
\name{Keisuke Kinoshita, Marc Delcroix, Shoko Araki, Tomohiro Nakatani}
\address{NTT Communication Science Labs, NTT corporation}

\begin{document}

%
\maketitle
\begin{abstract}
Automatic meeting analysis is an essential fundamental technology required to let, e.g. smart devices follow and respond to our conversations. 
To achieve an optimal automatic meeting analysis,  
we previously proposed an all-neural approach that jointly solves source separation, 
speaker diarization and source counting problems in an optimal way 
(in a sense that all the 3 tasks can be jointly optimized through error back-propagation).
It was shown that the method could well handle simulated clean (noiseless and anechoic) dialog-like data,
and achieved very good performance in comparison with several conventional methods.
However, it was not clear whether such all-neural approach would be successfully generalized to 
more complicated real meeting data containing more spontaneously-speaking speakers, 
severe noise and reverberation, and how it performs in comparison with the state-of-the-art systems in such scenarios.
In this paper, 
we first consider practical issues required for improving the robustness of the all-neural approach,
and then experimentally show that, even in real meeting scenarios, 
the all-neural approach can perform effective speech enhancement, 
and simultaneously outperform state-of-the-art systems.

\end{abstract}
\begin{keywords}
Source separation, source number counting, speaker tracking, diarization
\end{keywords}

\vspace{-0mm}
\section{Introduction}
\label{sec:intro}

Automatic meeting/spoken-conversation analysis is one of essential fundamental technologies required 
for realizing, e.g. communication robots that can follow and respond to our conversations. 
The meeting analysis comprises several tasks, namely (a) diarization, i.e.,  determining who is speaking when, (b) source counting, i.e., estimating the number of speakers in the conversation, 
(c) source separation, and (d) automatic speech recognition (ASR), i.e., transcribing the separated streams corresponding to each person.
While ideally these tasks should be jointly accomplished to realize optimal meeting analysis,
most studies focus on one of the aforementioned tasks,
since each task itself is already very challenging in general.

For example, a considerable number of research has been carried out for developing reliable diarization systems \cite{Diarization_review, DIHARD_data, AMI_data}.
Most of the conventional diarization approaches perform block or block-online processing with the following two steps \cite{Diarization_review, Araki_ICASSP2007, Araki_HSCMA_2008, DIHARD_BUT, DIHARD_JHU}.
First, at each block, they perform source separation (if necessary) and obtain speaker identity features concerning each speaker, 
in the form of, e.g. i-vector \cite{i-vector}, x-vector \cite{x-vector}, or spatial signature \cite{Araki_ICASSP2007, Araki_HSCMA_2008, Drude_ICASSP2018}.
Then, the correct association of speaker identity information among blocks, i.e., diarization results, is estimated by
clustering these features by using e.g. agglomerative hierarchical clustering \cite{DIHARD_JHU}.
Although these conventional algorithms can achieve reasonable diarization performance,
the results are not guaranteed to be optimal,
because the steps concerning speaker identity feature extraction and clustering are done independently.
Focusing on this limitation, \cite{Fujita_IS2019} recently proposed a neural network (NN)-based diarization approach 
that directly outputs diarization results (without any clustering stage), 
and showed that it can outperform a conventional two-stage approach \cite{DIHARD_JHU} in CALLHOME task 
where two people speak over a phone channel.
Note that, although some diarization systems can deal with overlapped speech, it does not mean that they can perform speech enhancement, i.e., separation and denoising, which is
eventually required for the meeting analysis.

Another key challenge for automatic meeting analysis is the separation of overlapped speech. 
Perhaps surprisingly, even in professional meetings, the percentage of overlapped speech, 
i.e., time segments where more than one person is speaking, 
is in the order of 5 to 10\%, 
while in informal get-togethers 
it can easily exceed 20\% \cite{onlineRSAN_ICASSP2019}.
To address the source separation problem, recently, many NN-based single-channel approaches have been proposed,
such as Deep Clustering (DC)\cite{Hershey_ICASSP16}, 
and Permutation Invariant Training (PIT)\cite{Yu2016, Kolbaek2017}.
DC can be viewed as two-stage algorithms, where in the first stage embedding vectors are estimated for each time-frequency (T-F) bin. 
In the second stage, these embedding vectors are clustered to obtain separation masks, given the correct number of clusters, i.e., sources.
PIT, on the contrary, is a single-stage algorithm, because it lets NNs directly estimate source separation masks.
In PIT, however, the NN architecture depends on the maximum number of sources to be extracted. 

To lift this constraint on the number of sources, 
we proposed Recurrent Selective Attention Network (RSAN) that is a purely NN-based mask estimator
capable of separating an arbitrary number of speakers and simultaneously counting the number of speakers in a mixture \cite{RSAN}.
It extracts one source at a time from a mixture and recursively performs this process until all sources are extracted.
The RSAN framework is based on a recurrent NN (RNN) 
which can learn and determine how many computational steps/iterations
have to be performed \cite{Graves_2016_arxiv_adaptiveRNN}.
Following the idea of the recursive source extraction, \cite{Takahashi_Interspeech2019} proposed to incorporate a time-domain audio separation network (TasNet) \cite{luo2019conv} into the RSAN framework.

To go one step further toward ideal meeting analysis that deals with aforementioned tasks (a)(b)(c) simultaneously,  
in \cite{onlineRSAN_ICASSP2019}, RSAN was extended to an all-neural block-online approach (hereafter, online RSAN) 
that simultaneously performs source separation, source number counting and even speaker tracking from a block to a block, 
i.e., performing the diarization-like process.
It was shown that online RSAN can handle well-controlled scenarios such as clean (noiseless and anechoic) simulated dialog-like data 
of 30 seconds, and outperform conventional systems in terms of source separation and diarization performance.

However, it was not clear from our past studies 
whether such all-neural approach, i.e., online RSAN, can be generalized 
to realistic meeting scenarios containing more spontaneously-speaking speakers, 
severe noise and reverberation, and how it performs in comparison with the state-of-the-art systems.
To this end, this paper focuses on the application of the online RSAN model 
to the realistic situations, and its evaluation in comparison with state-of-the-art diarization algorithms.
We first review the online RSAN model (in Section~\ref{sec:prop})
and introduces practical techniques to increase robustness against real meeting data,
such as a decoding scheme that mitigates over-estimation error in source counting (in Section~\ref{sec:decode}).
We then carry out experiments with real and simulated meeting data involving up to 6 speakers, 
containing a significant amount of noise and reverberation.
Then, finally, it will be shown that, even in such difficult scenarios, 
online RSAN can still perform effective speech enhancement, i.e., source separation and denoising, 
and simultaneously outperform a state-of-the-art diarization system \cite{dihard19} developed 
for a recent challenge \cite{DIHARD_data}.
This finding is the main contribution of this paper. 
Before concluding this paper, we also discuss the challenges that remain.

\section{Overall Structure of online RSAN}
\label{sec:prop}

\begin{figure}[t]
 \begin{center}
  \includegraphics[width=90mm]{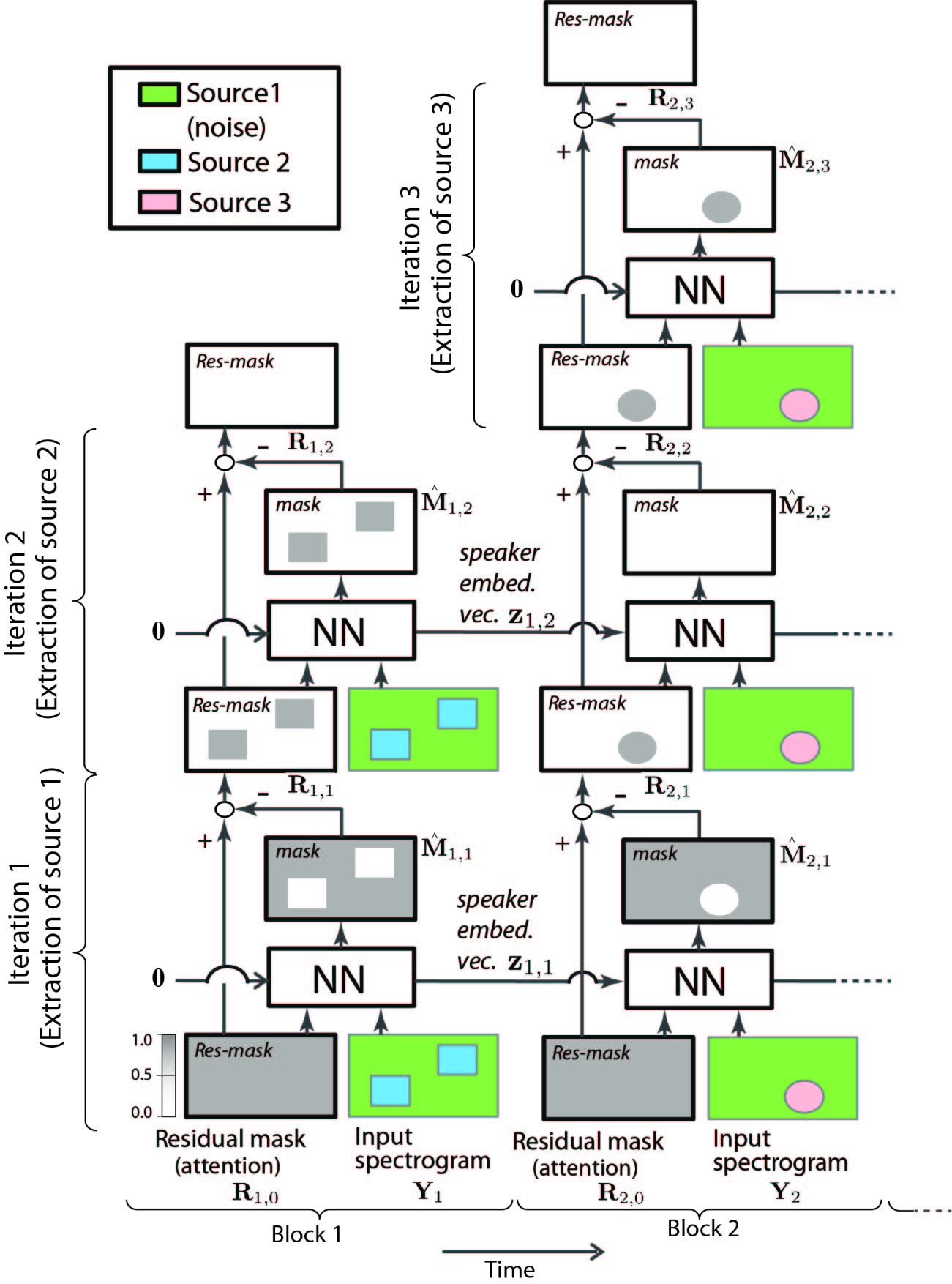}
     \end{center}
   \caption{Overview of online RSAN framework}
 \label{fig:overview}
\end{figure}

Figure \ref{fig:overview} summarizes how online RSAN works on the first 2 blocks of an example mixture containing three sources.
Since the model works in a block-online manner,
we first split the input magnitude spectrogram $\mat{Y}$ into $B$ consecutive blocks of equal time length, $[\mat{Y}_1,\ldots,\mat{Y}_b,\ldots,\mat{Y}_B]$, before feeding it to the system.

Online RSAN estimates a source extraction mask $\hat{\mathbf{M}}_{b,i}$ in each $b$-th block recursively to extract all source signals therein, 
while judging at each $i$-th source extraction iteration whether or not to proceed to the next iteration to extract more source signals.
The same neural network ``NN'' is repeatedly used for each block and iteration.
At every iteration in $b$-th block, NN receives three inputs, 
a residual mask $\mat{R}_{b,i}$ (res-mask in the figure), 
an input spectrogram $\mat{Y}_b$, 
and an auxiliary feature $\vect{z}_{b-1,i}$ (speaker embed.~vec. in the figure) 
to estimate a mask for a specific speaker $\hat{\mathbf{M}}_{b,i}$ 
and a speaker embedding vector representing that specific speaker $\vect{z}_{b,i}$ as:
\begin{equation}
    \hat{\mathbf{M}}_{b,i}, \vect{z}_{b,i} = \textrm{NN}(\mat{Y}_b, \mathbf{R}_{b,i}, \vect{z}_{b-1,i}).
\end{equation}
The residual mask can be seen as an attention map that informs the network
where to attend to extract a speaker that was not extracted in previous iterations in the current block.
At every first iteration in the $b$-th block, the residual mask is initialized with $\mat{R}_{b,0}=\mat{1}$.

In the first processing block, $b=1$, where we have two sources, online RSAN performs two source extraction iterations 
to extract all sources.
Since it is the first block, no speaker information is available from the previous block. 
Therefore, the input speaker information is set to zero, $\vect{z}_{0,i}=\vect{0}$. 
Without guidance, the network decides on its own in which order it extracts the source signals.
Fig.~\ref{fig:overview} shows a situation where source 1 is extracted at the first iteration.
Then, the first iteration is finished after generating another residual mask for the next iteration 
by subtracting the estimated source extraction mask from the input residual mask as $\mathbf{R}_{1,1} = \mathbf{R}_{1,0}-\hat{\mathbf{M}}_{1,1}$.
At the second iteration, the network receives the residual mask $\mathbf{R}_{1,1}$ 
and input spectrogram $\mathbf{Y}_1$ to estimate a mask for another source.
Then, it follows the same procedure as the first iteration.
The network decides to stop the iteration 
by examining how empty the updated residual mask is.
Specifically, the source extraction process is stopped in iteration $i$ if
$\frac{1}{TF} \sum_{tf} \mathbf{R}_{i,tf} < t_{\text{res-mask}}$ 
where $T$ and $F$ correspond to the total number of time frames and frequency bins in the block,
and $t_{\text{res-mask}}$ is an appropriate threshold.

Note that the speaker embedding vector $\vect{z}_{b,i}$ is passed to the next time block, $b+1$, 
and guides the $i$-th iteration on that block to extract the same speaker as in $(b,i)$. 
In the figure, it can be seen that the green source (source 1) is always extracted in the first, 
the blue in the second and the pink in the third iterations. 
If a source happens to be silent in a particular block (see the 2nd iteration in block 2), 
the estimated mask is to be filled with zeros ($\hat{\mat{M}}_{b,i} = \vect{0}$), 
and the residual mask is to stay unmodified ($\mat{R}_{b,i} = \mat{R}_{b,i-1}$).

At the second iteration of block 2, 
the criterion to stop the source extraction iteration is not met because the residual mask is not empty.
In such a case, the model increases the number of iterations to extract any new speaker
until the stopping criterion is finally met.
Overall, by having this structure, the model can perform jointly source separation, counting and speaker tracking.

To deal with background noise,
in this study, we force online RSAN to estimate a mask for the noise always at the first source extraction iteration 
in each block (see Fig. \ref{fig:overview}).
With this scheme, we can easily identify signals corresponding to background noise among separated streams 
and discard them if necessary.

\section{Summary of practical techniques \\ required to handle real meeting data}
\label{sec:practical}
This section summarizes techniques that we incorporated into online RSAN
to cope with noisy reverberant real meeting data.
They are divided into categories of input feature, training schemes, and decoding scheme
and summarized in the following dedicated subsections. 

\subsection{Input feature: Multichannel feature}
As in much past literature, e.g. \cite{Yoshioka_ICASSP2018},
our preliminary experiments confirmed that a multichannel feature as an additional input 
helps improve separation performance in reverberant environments.
In this study, therefore, the inter-microphone phase difference (IPD) feature proposed in \cite{Yoshioka_ICASSP2018} 
is concatenated by default to the magnitude spectrogram and used as input to online RSAN.

\subsection{Training scheme}
To train online RSAN,
we used training data comprising a pair of noisy reverberant meeting-like mixtures,
and corresponding noiseless reverberant single-speaker signals.

\subsubsection{Cost function for model training}
\label{sec:prop:training}
During training, the network is unrolled over multiple blocks and iterations and 
was trained with back-propagation using the following multi-task cost function:
\begin{equation}
\mathcal{L} = \Lmse + \alpha \Lresmask + \beta \Ltriplet
\end{equation}
In the following, we explain each term on the right-hand side of the above equation,
starting from $\Lmse$.

At each iteration, the network is required to output a mask for a certain source, but the order of source extraction when they first appear is not predictable.
In such a case, a permutation-invariant loss function is required.
Once a source was extracted and the permutation was chosen to minimize the error on its first appearance, 
its position in the iteration process is fixed for any following blocks, as the embedding vectors are passed and thus the desired output order is known.
Consequently, a {\it partially} permutation-invariant utterance-level mean square error (MSE) loss can be used for online RSAN as:
\begin{equation}
\Lmse = \frac{1}{IB} \sum_{i,b} |\hat{\mat{M}}_{i,b}\odot \mat{Y} - \mat{A}_{\phi_b}|^2,
\end{equation}
where $\mat{A}_{\phi_b}$ is a target reverberant single-speaker signal.
\footnote{To handle reverberant mixture, it was found in our preliminary investigation 
that the target signal $mat{A}_{\phi_b}$ has to be magnitude spectrum of reverberant (not anechoic clean) speech}.
When a source was active before, but is silent on the current block, a silent signal $\mat{A}_{b,i}=\mat{0}$ are inserted as a target.
The permutation $\phi_b$ for $b$-th block is formed by concatenating the permutation used for the last block $\phi_{b-1}$ with the permutation $\phi^*_b$ that minimizes the utterance-level separation error for the
newly discovered sources in block $b$.
To force the network to estimate a mask for the noise always at the first source extraction iteration,
we always inserted noise magnitude spectrogram as a target at the first iteration,
and from the second iteration, we used the above partially permutation-invariant loss.

$\Lresmask$ is a cost function related closely to the source counting performance.
To meet the speaker counting and iteration stopping criterion when all sources are extracted from a mixture, 
we minimize this cost and pushes the values of the residual mask to $0$ if no speaker is remaining (see \cite{RSAN} for more details).
we minimize the following $\Lresmask$ and pushes the values of the residual mask to $0$ if no speaker is remaining.
\begin{equation}
\Lresmask = \sum_{b,tf} \left[\max\left(1-\sum_{i}\hat{\mat{M}}_{b,i}, 0\right)\right]_{tf}
\end{equation}

$\Ltriplet$ is a triplet loss that is shown to help increase speaker discrimination capability,
by ensuring the cosine similarity between each pair of embedding vectors for the same speaker 
is greater than for any pair of vectors of differing speakers.
It is formed by first choosing one anchor vector $\vect{a}$, a positive vector $\vect{p}$ belonging to the same speaker as $\vect{a}$,
and a negative vector $\vect{n}$ belonging to a different speaker from $\vect{a}$, 
from a set of speaker embedding vectors within one minibatch.
Based on the cosine similarities between the anchor and negative vectors $s_i^{an}$,
and the anchor and the positive vectors $s_i^{ap}$, 
the triplet loss for $N$ triplets can be calculated as \cite{Li2017}: 
\begin{equation}
\Ltriplet = \sum_{i=0}^{N} \max(s_i^{an} - s_i^{ap} + \delta, 0).
\end{equation}
where $\delta$ is a small positive constant.
Interestingly, in this study, this loss was found to improve not only speaker discrimination capability
but also speaker tracking capability of online RSAN.
When training with meeting-like data where people speaks intermittently,
one minibatch is usually formed with only a part of a meeting, in which very often certain speaker speaks only one time
and remains silent to the end of this meeting excerpt (although he/she may start speaking again in a later part of the meeting).
If we do not use the triplet loss, the network is not encouraged to keep remembering such a person to the end of the meeting,
and eventually, it starts estimating a speaker embedding vector irrelevant to the speaker.
To circumvent such issue and make the network ready always for a situation when he/she starts speaking again,
we can use the triplet loss and make the network always output speaker embedding vectors that are consistent over frames
no matter whether the speaker is speaking or not.

\subsubsection{Noise mask}
To deal with background noise in the mixture, we let the network to estimate a mask for background noise 
always at the first source extraction iteration in each block.
To force such behavior to the network, we used the following permutation variant loss (as oppose to permutation
invariant loss) for the mask for background noise:
\begin{equation}
\Lmse = \frac{1}{B} |\hat{\mat{M}}_{1,b}\odot \mat{Y} - \mat{A}_b^{(\textrm{noise})}|^2,
\end{equation}
where $\mat{A}_b^{(\textrm{noise})}$ is the magnitude spectrum of background noise.
With this scheme, we can easily find and discard signals corresponding to background noise from enhancement results.

\subsubsection{Teacher forcing for iterative source extraction}
During training, when calculating residual mask $\mathbf{R}_{b,i+1}$ at $b$-th block for the $(i+1)$~-th iteration,
instead of using the estimated source extraction mask at the $i$-th iteration,
we can calculate it by using an oracle source extraction mask $\hat{\mathbf{M}}_{b,i}^{\textrm{(oracle)}}$ 
based on the estimated permutation~\cite{RSAN},
i.e., $\mathbf{R}_{b,i+1} = \mathbf{R}_{b,i}-\mathbf{M}_{b,i}^{\textrm{(oracle)}}$.
In RNN literature, this form of training is known as teacher forcing.
While the effectiveness of this scheme was not so clear in the previous study \cite{RSAN},
we found it mandatory when we cope with noise and many speakers, 
both of which are inevitable during training with meeting-like data.

\subsection{Decoding scheme: decoding with consistency check}
\label{sec:decode}
Real meeting data contains a lot of unexpected sound events that are hardly observed in the training data,
such as laughing sounds, a sudden change in tone of voice, coughing sound, 
and rustling sounds from e.g. papers, to name a few.
All of these sounds can be a cause for online RSAN to mistakenly detect a new spurious speaker and increase the number of source count.
When it increases the source count by mistake because of such unexpected unseen sounds, 
it tends to wrongly split a speaker into two; the new embedding vector 
and an embedding vector already associated with the speaker.
It causes over-estimation errors in the source counting,
and degrades embedding vector representation of the speaker, leading to degradation in overall performance.

Let us denote the speaker embedding vector set
at $b'$-th block as $\{\vect{z}_{b',i}\}_{1 \leq i \leq I_{b'}}$
where $I_{b'}$ corresponds to the total number of iteration in $b'$-th block.
Then, to reduce such over-estimation error in source counting, 
we propose to perform the following consistency check for the speaker embedding vector set, 
$\{\vect{z}_{b',i}\}_{1 \leq i \leq I_{b'}}$, when online RSAN increases the speaker count.
Specifically, we propose to decode all the past blocks with the embedding vector set $\{\vect{z}_{b',i}\}_{1 \leq i \leq I_{b'}}$.
Then, if masks estimated with $\vect{z}_{b',I_{b'}}$, $\{\hat{\mat{M}}_{b,I_{b'}}\}_{1 \leq b \leq b'}$, does not exceed $t_{\text{res-mask}}$,
it indicates that a speaker corresponding to $\vect{z}_{b',I_{b'}}$ did not indeed appear 
in the past blocks and appeared for the first time at $b'$-th block. 
And thus, we accept the increase in the source count and keep using $\{\vect{z}_{b',i}\}_{1 \leq i \leq I_{b'}}$ for further process.
Otherwise, we do not accept the increase in the speaker count,
and discard $\{\vect{z}_{b',i}\}_{1 \leq i \leq I_{b'}}$ and replace the set of embedding vectors 
with ones from the previous block $\{\vect{z}_{b'-1,i}\}_{1 \leq i \leq I_{b'-1}}$ for further process.

\section{Experiments}
\label{sec:exp}
In this section, we evaluate online RSAN in comparison with state-of-the-art diarization methods,
and shows its effectiveness. 

\subsection{Experimental conditions}
\subsubsection{Data}
We generated three sets of noisy reverberant multi-speaker datasets for training;
(dataset A) 20000 mixtures each of which is 10 seconds in length, and contains 1 or 2 speakers' speech signals, 
(dataset B) 10000 mixtures each of which is 60 seconds in length, and contains 1 to 6 speakers' speech signals,
and (dataset C) 2372 mixtures each of which is 60 seconds in length, and contains 1 to 6 speakers' speech signals. 
To all dataset, we added CHiME4 noise with SNR of 10 to 20~dB, and reverberation of $\textrm{RT}_{\textrm{60}}$ of 300 to 700~ms.
Utterances for dataset A and B are taken from WSJ0 \cite{WSJ0}, i.e., read speech, 
while those for dataset C are taken from headset recordings of real meetings recorded in our office, i.e., spontaneous speech. 
In dataset A, each mixture was generated such that the first \SI{5}{\second} 
contain one or two speakers with a probability of \SI{50}{\percent} each,
while the second half contains zero, one  or two speakers 
with a probability of \SI{15}{\percent}, \SI{55}{\percent} and \SI{30}{\percent}, respectively.
Similarly, in dataset B, 
the first \SI{5}{s} of the test utterance contains zero or one speaker with a probability of \SI{50}{\percent} each,
while the mixture in the remaining time is generated such that it contains zero, 
one, two or three speakers with a probability of \SI{5}{\percent}, \SI{75}{\percent} and \SI{15}{\percent} and \SI{5}{\percent} respectively.


Evaluation was done with two datasets; (1) simulated meeting-like data
comprising 1000 mixtures similar to dataset B but with unseen speakers,
and (2) real meeting data recorded at our office with a distant microphone-array \cite{ArakiHSCMA2017}.
Real meeting dataset consists of 8 meetings, each of which is 15 to 20 minutes in length.
The number of meeting participants varies from 4 to 6, all of who are unseen during training.
The meeting recording contains a significant amount of babble noise (SNR of 3 to 15~dB),
and reverberation of $\textrm{RT}_{\textrm{60}}$ of 500~ms.
The percentage of overlapped speech in these meetings is found to be $25.7$~\% on average.

\subsubsection{Implementation details of online RSAN}
NN architecture and hyper-parameters for online RSAN was same as \cite{onlineRSAN_ICASSP2019}.
It consists of one fully connected layer on top of two BLSTM layers.
Multichannel input feature was calculated based on signals observed at 2 microphones,
and thus overall online RSAN model in this study is a 2-channel system.
For the evaluation based on the simulated meeting-like data, 
the online RSAN model was first trained with the training dataset A for 300 epochs,
and then further trained with dataset B for 50 epochs.
Then, to cope with real meeting data, the model was further trained with dataset C for 5 epochs.
The block size of online RSAN was set at 10 seconds.
$t_{\text{res-mask}}$ was set at 0.2.
To obtain diarization results with online RSAN,
we performed power-based voice activity detection (VAD) on extracted streams 
based on a threshold value common to one meeting.

\subsubsection{Methods to be compared with}
In the evaluation with the simulated meeting-like data, 
the performance of online RSAN was compared with a system similar to a top-performing system \cite{DIHARD_JHU} in DIHARD-1 challenge \cite{DIHARD_data}.
For this, we used off-the-shelf implementation and model from \cite{dihard19}.
Since it is based on clustering of x-vectors \cite{x-vector},
it will be referred to as ``x-vector clustering" hereafter.
It is a single-channel system.

For the real meeting data evaluation, the performance of online RSAN was compared with ``x-vector clustering"
and a multi-channel diarization method based on online clustering of Time-Difference-Of-Arrival (TDOA) feature \cite{Hori_TASLP2011},
which will be referred to as ``TDOA clustering". 
The TDOA feature was calculated based on signals observed at 8 microphones.
Diarization performance was evaluated in terms of diarization error rate (DER) \cite{der} including speaker overlapped segments,
while the speech enhancement performance was evaluated in terms of signal-to-distortion ratio (SDR) in BSSeval \cite{BSSeval}.
The sampling frequency was 8k~Hz for all the methods.

\subsection{Experiment 1: Evaluation with simulated meeting-like data}
Before proceeding to evaluation with real meeting data,
we briefly examine the performance of online RSAN and whether it is ever possible to
cope with noisy reverberant mixtures containing many speakers.
Table \ref{tbl:results_sim1} shows DERs of online RSAN and x-vector clustering, averaged over 1000 mixtures.
It was found that online RSAN works for noisy reverberant data, and outperformed the state-of-the-art x-vector clustering.
SDR improvement obtained with online RSAN was 10.01~dB, which we believe is reasonably high.

\begin{table}[t]
 \centering
  \caption{DERs for simulated meeting-like data (\%)}
 \label{tbl:results_sim1}
\begin{tabular}{|c|c|c|c|}
\hline
\begin{tabular}{c} x-vector clustering \end{tabular} & \begin{tabular}{c} Online RSAN \end{tabular}  \\
\hline
44.39 & \textbf{33.7}     \\  \hline
\end{tabular}
\end{table}

\begin{figure*}[t]
 \begin{center}
  \includegraphics[width=130mm]{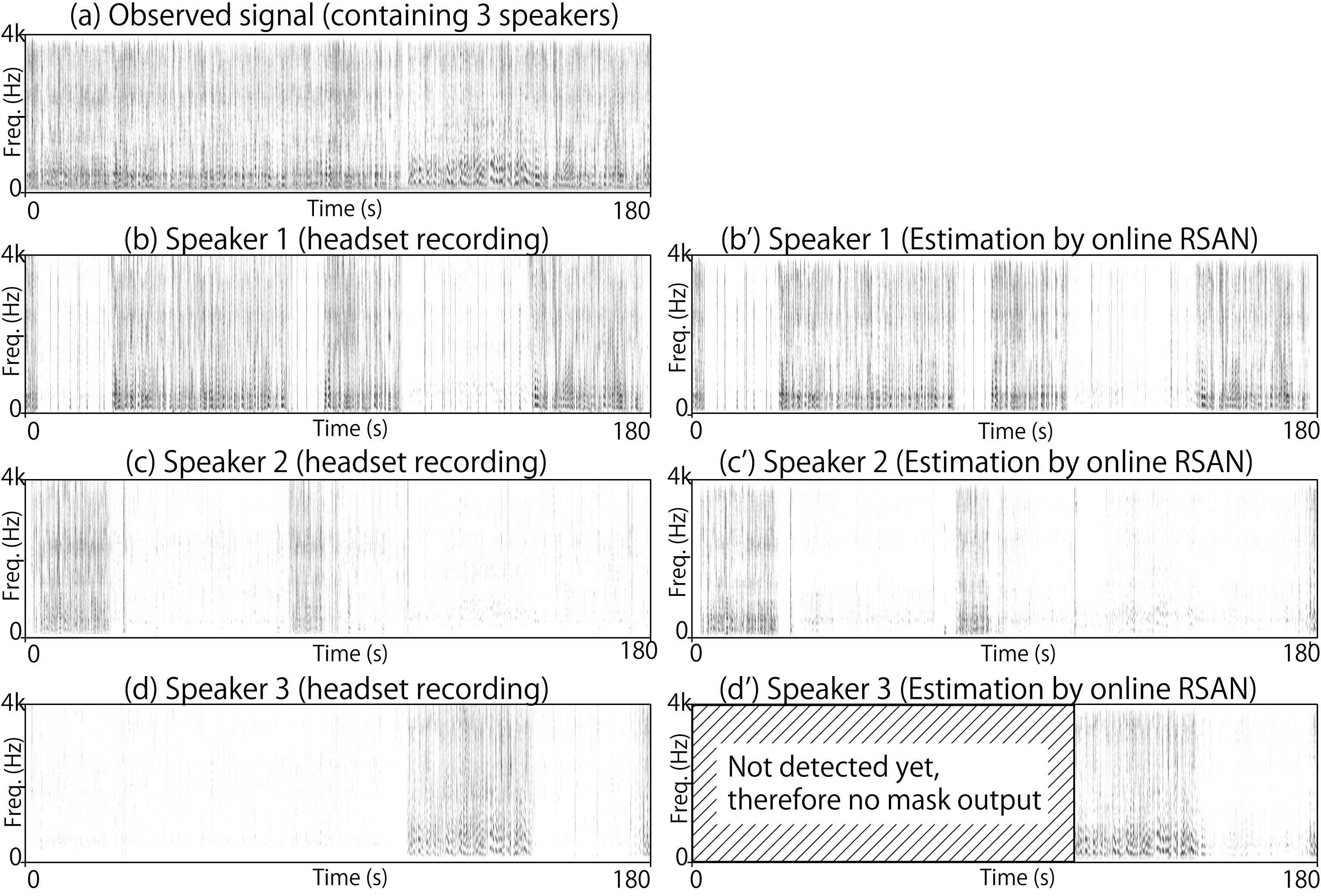}
     \end{center}
   \caption{Spectrograms of 3 minute excerpt from real meeting data: observed signal (a), headset recordings of each speaker (b,c,d), and signals estimated by online RSAN (b',c',d')}
 \label{fig:spectrogram}
\end{figure*}

\subsection{Experiment 2: Real meeting data}
Now we evaluate the performance of online RSAN with real data.
We first evaluate effect of consistency-checking decoding (proposed in section \ref{sec:decode}).
Also, since it may not be clear how much and what kind of training data would be necessary 
for online RSAN to cope with real data,
we examined the effect of training with dataset C, i.e., spontaneous speech.
Table \ref{tbl:results_real1} suggests that both training with spontaneous speech 
and decoding with the consistency check is important. 
Note that, although one may think that DERs here are very high, 
it is in a similar range as DIHARD2 as you can see in \cite{dihard19}.

\begin{table}[t]
\centering
 \caption{Effect of model training with spontaneous speech, and decoding with consistency check when dealing with real meeting data}
  \label{tbl:results_real1}
   \begin{tabular}{l c}
    \hline
    system & DER (\%) \\ \hline
    Online RSAN trained only with dataset A \& B       & 73.9 \\
    \ \ + dataset C                                &  56.3 \\
    \ \ \ \ +decoding with consistency checking   &  \textbf{49.6} \\ \hline
  \end{tabular}
\end{table}

Now, let us compare, with the other methods, the performance of online RSAN model trained 
on dataset A, B, and C
and decoded with the consistency check.
While we use a different threshold for power-based VAD for each meeting for table \ref{tbl:results_real1}
to take a closer look at the difference between each setting,
we reran the experiments with a common threshold for all meetings 
to make a fair comparison with the other methods.
The threshold was determined based on the validation dataset.
Table \ref{tbl:results_each} shows DERs of online RSAN, 
x-vector clustering and TDOA clustering.
While there are some sessions where deep learning-based algorithms, i.e. x-vector clustering and online-RSAN did not work well,
on average, even in the real meeting scenario, online RSAN can outperform these conventional approaches.

Figure \ref{fig:spectrogram} shows spectrograms of unprocessed real meeting data 
(3-minute excerpt from one of the meetings used for the evaluation), 
corresponding headset recording of each speaker, and signals estimated by online RSAN. 
Noise signals estimated by online RSAN are omitted from the figure.
Audio examples corresponding to the spectrograms are available on our web-page \cite{demo_page}.
As it can be seen, the headset recordings and the estimated signals look quite similar,
which suggests that online RSAN extracted each speaker's voice clearly, and counted the number of speakers correctly. 
Note that, as the third speaker did not speak for the first 2 minutes, online RSAN did not output masks for that speaker in that period, 
and once that speaker starts speaking, it correctly increased the number of source extraction iteration to 4 (i.e., 3 speakers + noise) 
and started tracking that speaker.


\begin{table*}[]
\centering
\label{tbl:results_each}
\caption{DERs for noisy reverberant real meetings for each system}
\begin{tabular}{|c|c|c|c|c|}
\hline
\begin{tabular}{c} Meeting \\ ID \end{tabular} & \# of spk. & \begin{tabular}{c} x-vector \\ clustering \end{tabular} & \begin{tabular}{c} TDOA \\ clustering \end{tabular} & \begin{tabular}{c} Online \\ RSAN \end{tabular}  \\ \hline
1 & 6 &  51.2 & 46.8          & \textbf{41.4}    \\ \hline
2 & 6 &  61.8 & 64.6          & \textbf{58.4}    \\ \hline
3 & 6 &  73.5 & 62.6          & \textbf{49.0}   \\ \hline
4 & 5 &  57.6 & \textbf{23.8} & 55.7            \\ \hline
5 & 5 &  64.2 & \textbf{47.5} & 72.5             \\ \hline
6 & 6 &  71.4 & 67.2          & \textbf{40.3}     \\ \hline
7 & 4 &  68.1 & 73.6          & \textbf{45.5}     \\ \hline
8 & 4 &  72.4 & 70.9          & \textbf{48.8}     \\ \hline \hline
  & Average  &  65.0 & 57.1   & \textbf{51.4}     \\  \hline
\end{tabular}
\end{table*}

\subsection{Discussion}
Although online RSAN could cope with real meeting data to some extent, 
its performance was far from ``perfect".
We found that most errors in the online RSAN process are due to its insufficient performance in source separation and tracking.
It sometimes confidently extracts or tracks two different speakers' signals 
with one speaker embedding vector $\{\vect{z}_{b,i}\}_{1 \leq b \leq B}$,
probably because their voice characteristics are similar from the system's point of view.
This type of error should be reduced by, for example, 
employing more advanced NN architecture \cite{Takahashi_Interspeech2019}, 
and increasing the number of speakers in training data like we propose in \cite{Delcroix_ICASSP2020}.

\section{Conclusion}
This paper proposed several practical techniques required for all-neural diarization, source separation and 
source counting model called online RSAN to cope with real meeting data.
It was shown that incorporation of the proposed consistency-checking decoding 
and training with spontaneous speech is effective.
Based on the experiments with real meeting recordings, 
online RSAN was shown to perform effective speech enhancement, and simultaneously outperform state-of-the-art diarization systems.
Our future work includes incorporation of advanced source separation NNs into online RSAN,
and evaluation in terms of ASR accuracy.


\bibliographystyle{./bibliography/IEEEbib}
\bibliography{./bibliography/refs}

\begin{thebibliography}{10}

\bibitem{Diarization_review}
X.~Anguera, S.~Bozonnet, N.~Evans, C.~Fredouille, G.~Friedland, and O.~Vinyals,
\newblock ``Speaker diarization: A review of recent research,''
\newblock {\em IEEE Transactions on Audio, Speech, and Language Processing},
  vol. 20, no. 2, pp. 356--370, Feb 2012.

\bibitem{DIHARD_data}
N.~Ryant, K.~Church, C.~Cieri, A.~Cristia, J.~Du, S.~Ganapathy, and
  M.~Liberman,
\newblock {\em First {DIHARD} Challenge Evaluation Plan}, 2018,
\newblock {https://zenodo.org/record/1199638}.

\bibitem{AMI_data}
J.~Carletta, S.~Ashby, S.~Bourban, M.~Flynn, M.~Guillemot, T.~Hain, J.~Kadlec,
  V.~Karaiskos, W.~Kraaij, M.~Kronenthal, G.~Lathoud, M.~Lincoln, A.~Lisowska,
  I.~McCowan, W.~Post, D.~Reidsma, , and P.~Wellner,
\newblock ``The {AMI} meeting corpus: A pre-announcement,''
\newblock in {\em The Second International Conference on Machine Learning for
  Multimodal Interaction, ser. MLMI'05}, 2006, pp. 28--39.

\bibitem{Araki_ICASSP2007}
S.~Araki, H.~Sawada, and S.~Makino,
\newblock ``Blind speech separation in a meeting situation with maximum {SNR}
  beamformers,''
\newblock in {\em Proc. 2007 IEEE International Conference on Acoustics, Speech
  and Signal Processing (ICASSP)}, April 2007, vol.~1, pp. I--41--I--44.

\bibitem{Araki_HSCMA_2008}
S.~Araki, M.~Fujimoto, K.~Ishizuka, H.~Sawada, and S.~Makino,
\newblock ``A {DOA} based speaker diarization system for real meetings,''
\newblock in {\em Proc. Hands-Free Speech Communication and Microphone Arrays},
  2008, pp. 29--32.

\bibitem{DIHARD_BUT}
M.~Diez, F.~Landini, L.~Burget, J.~Rohdin, A.~Silnova, K.~Zmolikova,
  O.~Novotn{\'y}, K.~Vesel{\'y}, O.~Glembek, O.~Plchot, L.~Mo{\v s}ner, and
  P.~Mat{\v e}jka,
\newblock ``{BUT} system for {DIHARD} speech diarization challenge 2018,''
\newblock in {\em Proc. Interspeech 2018}, 2018, pp. 2798--2802.

\bibitem{DIHARD_JHU}
G.~Sell, D.~Snyder, A.~McCree, D.~Garcia-Romero, J.~Villalba, M.~Maciejewski,
  V.~Manohar, N.~Dehak, D.~Povey, S.~Watanabe, and S.~Khudanpur,
\newblock ``Diarization is hard: Some experiences and lessons learned for the
  {JHU} team in the inaugural {DIHARD} challenge,''
\newblock in {\em Proc. Interspeech 2018}, 2018, pp. 2808--2812.

\bibitem{i-vector}
N.~Dehak, P.~Kenny, R.~Dehak, P.~Dumouchel, , and P.~Ouellet,
\newblock ``Front-end factor analysis for speaker verification,''
\newblock {\em IEEE Trans. Audio, Speech, and Language Processing}, vol. 19(4),
  pp. 788--798, 2011.

\bibitem{x-vector}
D.~Snyder, P.~Ghahremani, D.~Povey, D.~Garcia-Romero, Y.~Carmiel, , and
  S.~Khudanpur,
\newblock ``Deep neural network-based speaker embeddings for end-to-end speaker
  verification,''
\newblock in {\em Proc. IEEE Spoken Language Technology Workshop}, 2016.

\bibitem{Drude_ICASSP2018}
L.~Drude, T.~Higuchi, K.~Kinoshita, T.~Nakatani, and R.~Haeb-Umbach,
\newblock ``Dual frequency- and block-permutation alignment for deep learning
  based block-online blind source separation,''
\newblock in {\em Proc. 2018 IEEE International Conference on Acoustics, Speech
  and Signal Processing (ICASSP)}, 2018, pp. 691--695.

\bibitem{Fujita_IS2019}
Y.~Fujita, N.~Kanda, S.~Horiguchi, K.~Nagamatsu, and S.~Watanabe,
\newblock ``End-to-end neural speaker diarization with permutation-free
  objectives,''
\newblock in {\em Proc. Interspeech 2019}, 2019, pp. 4300--4304.

\bibitem{onlineRSAN_ICASSP2019}
T.~von Neumann and S.~Araki T. Nakatani R. Haeb-Umbach K.~Kinoshita,
  M.~Delcroix,
\newblock ``All-neural online source separation, counting, and diarization for
  meeting analysis,''
\newblock in {\em Proc. 2018 IEEE International Conference on Acoustics, Speech
  and Signal Processing (ICASSP)}, May 2019, pp. 91--95.

\bibitem{Hershey_ICASSP16}
J.~Hershey, Z.~Chen, J.~Le Roux, and S.~Watanabe,
\newblock ``Deep clustering: Discriminative embeddings for segmentation and
  separation,''
\newblock in {\em Proc. 2016 IEEE International Conference on Acoustics, Speech
  and Signal Processing (ICASSP)}, 2016, pp. 31--35.

\bibitem{Yu2016}
D.~Yu, M.~Kolb{\ae}k, Z.~Tan, and J.~Jensen,
\newblock ``Permutation invariant training of deep models for
  speaker-independent multi-talker speech separation,''
\newblock in {\em Proc. 2017 IEEE International Conference on Acoustics, Speech
  and Signal Processing (ICASSP)}, March 2017, pp. 241--245.

\bibitem{Kolbaek2017}
M.~Kolb{\ae}k, D.~Yu, Z.~Tan, and J.~Jensen,
\newblock ``Multitalker speech separation with utterance-level permutation
  invariant training of deep recurrent neural networks,''
\newblock {\em IEEE/ACM Transactions on Audio, Speech, and Language
  Processing}, vol. 25, no. 10, pp. 1901--1913, Oct 2017.

\bibitem{RSAN}
K.~Kinoshita, L.~Drude, M.~Delcroix, and T.~Nakatani,
\newblock ``Listening to each speaker one by one with recurrent selective
  hearing networks,''
\newblock in {\em Proc. 2018 IEEE International Conference on Acoustics, Speech
  and Signal Processing (ICASSP)}, April 2018, pp. 5064--5068.

\bibitem{Graves_2016_arxiv_adaptiveRNN}
A.~Graves,
\newblock ``Adaptive computation time for recurrent neural networks,'' 2016,
\newblock arXiv:1603.08983.

\bibitem{Takahashi_Interspeech2019}
N.~Takahashi, S.~Parthasaarathy, N.~Goswami, and Y.~Mitsufuji,
\newblock ``Recursive speech separation for unknown number of speakers,''
\newblock in {\em Proc. Interspeech 2019}, 2019, pp. 1348--1352.

\bibitem{luo2019conv}
Yi~Luo and Nima Mesgarani,
\newblock ``{Conv-TasNet}: Surpassing ideal time--frequency magnitude masking
  for speech separation,''
\newblock {\em IEEE/ACM Transactions on Audio, Speech, and Language Processing
  (TASLP)}, vol. 27, no. 8, pp. 1256--1266, 2019.

\bibitem{dihard19}
{{https://github.com/iiscleap/DIHARD-2019-baseline}}.

\bibitem{Yoshioka_ICASSP2018}
T.~Yoshioka, H.~Erdogan, Z.~Chen, and F.~Alleva,
\newblock ``Multi-microphone neural speech separation for far-field
  multi-talker speech recognition,''
\newblock in {\em Proc. 2018 IEEE International Conference on Acoustics, Speech
  and Signal Processing (ICASSP)}, May 2018, pp. 5739--5743.

\bibitem{Li2017}
C.~Li, X.~Ma, B.~Jiang, X.~Li, X.~Zhang, X.~Liu, Y.~Cao, A.~Kannan, and Z.~Zhu,
\newblock ``{Deep Speaker}: an end-to-end neural speaker embedding system,''
  2017,
\newblock arXiv:1705.02304v1.

\bibitem{WSJ0}
J.~Garofolo, D.~Graff, P.~Doug, and D.~Pallett,
\newblock {\em {CSR-I (WSJ0) Complete LDC93s6a}},
\newblock Linguistic Data Consortium, Philadelphia, New Jersey, 1993.

\bibitem{ArakiHSCMA2017}
S.~{Araki} et~al.,
\newblock ``Online meeting recognition in noisy environments with
  time-frequency mask based mvdr beamforming,''
\newblock in {\em 2017 Hands-free Speech Communications and Microphone Arrays
  (HSCMA)}, March 2017, pp. 16--20.

\bibitem{Hori_TASLP2011}
T.~{Hori}, S.~{Araki}, T.~{Yoshioka}, M.~{Fujimoto}, S.~{Watanabe}, T.~{Oba},
  A.~{Ogawa}, K.~{Otsuka}, D.~{Mikami}, K.~{Kinoshita}, T.~{Nakatani},
  A.~{Nakamura}, and J.~{Yamato},
\newblock ``Low-latency real-time meeting recognition and understanding using
  distant microphones and omni-directional camera,''
\newblock {\em IEEE Transactions on Audio, Speech, and Language Processing},
  vol. 20, no. 2, pp. 499--513, Feb 2012.

\bibitem{der}
NIST~Speech Group,
\newblock ``Spring 2007 (rt-07) rich transcription meeting recognition
  evaluation plan,'' 2007.

\bibitem{BSSeval}
E.~Vincent, R.~Gribonval, and C.~Fevotte,
\newblock ``Performance measurement in blind audio source separation,''
\newblock {\em IEEE Trans. Audio, Speech and Lang. Process.}, vol. 14, pp.
  1462--1469, 2006.

\bibitem{demo_page}
{{http://www.kecl.ntt.co.jp/icl/signal/kinoshita/publications/\\ICASSP20/onlineRSAN/index.html}}.

\bibitem{Delcroix_ICASSP2020}
M.~Delcroix, T.~Ochiai, K.~Zmolikova, K.~Kinoshita, N.~Tawara, T.~Nakatani, and
  S.~Araki,
\newblock ``Improving speaker discrimination of target speech extraction with
  time-domain speakerbeam,''
\newblock in {\em Proc. 2020 IEEE International Conference on Acoustics, Speech
  and Signal Processing (ICASSP) (submitted)}, 2020.

\end{thebibliography}

\end{document}